\documentclass[traditabstract]{aa} 
\usepackage{graphicx}
\usepackage{txfonts}
\begin{document}
   \title{The spectroscopic  evolution of the symbiotic-like recurrent nova V407 Cygni during its 2010 outburst}
   \subtitle{I. The shock and its evolution}

   \author{S. N. Shore\inst{1,2}, G. M. Wahlgren\inst{3,4}, T. Augusteijn\inst{5}, T. Liimets\inst{5,6}, K. L. Page\inst{7}  \\
   J. P. Osborne\inst{7},  A. P. Beardmore\inst{7}, P. Koubsky\inst{8}, M. \v{S}lechta\inst{8}, V. Votruba\inst{8}    }
        \institute{
   Dipartimento di Fisica ``Enrico Fermi'', Universit\`a di Pisa, largo B. Pontecorvo 3, I-56127 Pisa, Italy\\ \email{shore@df.unipi.it}
   \and
   INFN - Sezione di Pisa
   \and  
   The Catholic University of America, Dept. of Physics, 620 Michigan Ave NE, Washington DC, 20064, USA\\   \email{glenn.m.wahlgren@nasa.gov}
      \and
NASA-GSFC, Code 667, Greenbelt, MD, 20771, USA
\and
Nordic Optical Telescope, Apartado 474, E-38700 Santa Cruz de La Palma, Santa Cruz de Tenerife,
Spain\\ \email{tau@not.iac.es,tiina@not.iac.es}
\and
Tartu Observatory,  T\~oravere, 61602, Estonia\\ \email{sinope@aai.ee}
 \and
  Dept. of Physics and Astronomy, University of Leicester, Leicester LE17RH, UK\\\email{apb@star.le.ac.uk, julo@star.le.ac.uk,kpa@star.le.ac.uk}
   \and
  Astronomical Institute, Academy of Sciences of the Czech Republic,  CZ-251 65   Ond\v{r}ejov, Czech Republic\\ \email{koubsky@sunstel.asu.cas.cz}      
  }
     
              \date{Received ---; accepted ---}
 \abstract{V407 Cyg was, before 2010 Mar., known only as a D-type symbiotic binary system in which the Mira variable has a pulsation period of approximately 750 days, one of the longest known.  On 2010 Mar 10, it was discovered in outburst, eventually reaching $V$ $<$ 8.  This is the first recorded nova event for this system, but it closely resembles the spectroscopic development of RS Oph, the prototypical symbiotic-like recurrent nova.  It was also detected by $Fermi$ above 100 MeV and displayed strong, likely nonthermal centimeter wavelength radio emission.  
   Unlike classical novae occurring in compact cataclysmic binary systems, for which the ejecta undergo free ballistic expansion, this explosion occurred within the dense, complex wind of a Mira variable companion.  This paper concentrates on the development of the shock and its passage through the Mira wind.  We also present some constraints on the binary system properties.  
   Using medium and high resolution ground-based optical spectra, visual and $Swift$ UV photometry, and $Swift$ X-ray spectrophotometry, we describe the behavior of the high-velocity profile evolution for this nova during its first three months.  
  Using the diffuse interstellar bands visible in the high-resolution optical spectra, we obtain an extinction $E(B-V)$ $\approx$0.45$\pm$0.05.  The spectral type of the red giant during this period, when the star was at  $R$ minimum, was no earlier than  M7 III.
   The peak of the X-ray emission occurred at about day 40 with a broad maximum and decline after day 50.  The main changes in the optical spectrum began at around that time.  The He II 4686\AA\ line first appeared between days 7 and 14 and initially displayed a broad, symmetric profile that is characteristic of all species before day 60.   The profile development thereafter depended on ionization state.  Low-excitation lines remained comparatively narrow, with $v_{\rm rad,max}$ of order 200-400 km s$^{-1}$.  They were systematically more symmetric than lines such as [Ca V], [Fe VII], [Fe X], and He II, all of which showed a sequence of profile changes  going from symmetric to a blue wing similar to that of the low ionization species but with a red wing extended to as high as 600 km s$^{-1}$.   The [O I] 6300, 6364 doublet showed a narrow wind-emission component near the rest velocity of the system and a broad component, 200-300 km s$^{-1}$,  whose relative intensity increased in time. Forbidden lines of N II and O III had two separate contributors to the profiles, a broad line that increased in strength and velocity width, exceeding 200 km s$^{-1}$, and narrow components from a surrounding ionized region at higher velocity than the Mira wind.  The Na I D doublet developed a broad component with similar velocity width to the other low-ionization species.  The O VI Raman features observed in recent outbursts of RS Oph were not detected.   We interpret these variations as aspherical expansion of the ejecta within the Mira wind.  The blue side is from the shock penetrating into the wind while the red wing is from the low-density periphery.  The maximum radial velocities obey power laws, $v_{\rm max} \sim t^{-n}$ with n$\approx$1/3 for red wing and $\approx$0.8 for the blue. }

   \keywords{Stars-individual(V407 Cyg, RS Oph), symbiotic stars, physical processes, novae  }

  \titlerunning{The 2010 outburst of the recurrent nova V407 Cyg. I}   \authorrunning{S. N. Shore et al.}
   \maketitle

\section{Introduction}

The recurrent novae (RNe), the rarest and most extreme of the classical novae, are known in
two distinct varieties (for a general series of overviews and specific details on 
observations and models see Bode \& Evans 2008; Evans, Bode, O'Brien \& M.J. Darnley 2008; Bode 2010).  
Both show repeated episodes of explosive mass ejection, between $10^{-7}$M$_\odot$ and $10^{-6}$M$_\odot$, 
on recurrence timescales of decades with velocities (based on the ultraviolet resonance transitions) 
that can exceed 5000 km s$^{-1}$.  One group is indistinguishable from classical novae:  
compact systems with orbital period $< $1 day and a low-mass (normally main-sequence) star  
that transfers mass to its companion white dwarf (WD) by Roche lobe overflow in a stream and 
accretion disk (Shaefer 2010).  The other even rarer group, far less well studied but no less significant, 
resembles the symbiotic binaries, the so-called ``symbiotic-like recurrent novae" (SyRNe), 
in which the mass-losing star is a red giant (RG) (e.g. Gonzalez-Riestra 1992; Anupama \& Sethi 1994; 
Shore et al. 1996; Shore 2008; Anupama 2008; Schaefer 2010).  Until this year, there were only four known in the 
Galaxy: V745 Sco, V3890 Sgr, RS Oph, and T CrB.   Unlike the compact systems (e.g. U Sco), which 
have orbital periods coincident with classical cataclysmics, these systems have long orbital periods 
(years) and the WD likely accretes mass from the RG wind,  instead of a stream.  T CrB and 
RS Oph, for which accurate orbital parameters are known, host a WD whose mass exceeds that of the
RG and for which the degenerate is more massive than in the symbiotic novae (not to be confused, 
these do not explosively eject mass and, instead, display a thermonuclear runaway that enters a 
long-term period of burning with an inflated WD envelope).  The latter type are on the more active 
end of the symbiotic systems, the WD is the less massive component and $\le$1 M$_\odot$.   

Recurrent novae are produced by the initiation of a thermonuclear runaway provoked by accretion on massive WDs,  which are nearer to the
Chandrasekhar limit (for RS Oph, M$_{WD} \approx 1.2$ M$_\odot$) than for normal 
symbiotics or symbiotic novae ($<$ 1 M$_\odot$) (see Starrfield 2008; Anupama 2008, Mikolajewska 2008,  
Brandi et al. 2009).  The connection of these systems, and their supersoft XR 
emission after outburst, to SN Ia progenitors has long been a scenario (e.g. Munari \& Renzini 1992; 
Hernanz \& Jos\'e 2008; Sokoloski et al. 2006;  Walder et al. 2008; Bode 2010; 
Schaefer 2010).  The amount of mass that remains on the WD after the explosion is the critical 
determining factor. It is still unknown if there is a wind subsequent to the explosion that continues to 
remove the accreted mass, thus preventing the eventual critical accumulation to produce a collapse 
of the WD, or if the explosion ends abruptly and ejects only a fraction of the mass.  However, no 
symbiotic had ever been observed to display a classical nova explosion until that of V407 Cyg.  
This is why the system is both unique and significant. 

V407 Cyg, a D-type symbiotic binary with a long period Mira RG (Munari, Margoni, \& Stagni 1990), 
underwent its first ever recorded explosive mass ejection on 2010 Mar 10, reaching an unfiltered magnitude 
of 6.8 (Nishiyama \& Kabashima 2010) and $V$ magnitude peak of 7.9. Historically, V407 Cyg was first 
identified by its nova-like behavior in 1936 (Hoffmeister 1949, Meinunger 1966) in the $m_{pg}$ range 
14 to 16.5, but with identified contributions to the light curve from the Mira variable.
Since that time, it has been observed to periodically increase 
in brightness according to the long-period Mira, and minor, slowly developing  outbursts, never 
becoming brighter than approximately $V$ = 11.5 during the years 1984 - 1995 (Kolotilov et al. 1998). 
A slow-developing outburst in the late 1990s was identified from the light curve (Kolotilov et al. 2003, 
Shugarov et al. 2007), as it was claimed to become brighter than the  
outburst of 1936, but never being brighter than $V$ $\sim$ 11.8.  Large gaps in the light curve
exist from the discovery data epoch until around 1990, which are of sufficient duration to have missed any 
outburst similar to the current event.   Unfiltered optical monitoring of V407 Cyg during the two years
prior to the March outburst shows a step-like brightness
increase of nearly three magnitudes over a one year interval prior
to outburst (Abdo et al. 2010). The 2010 outburst increased
the brightness by an additional 2.5 magnitudes. These step-like
brightness increases, along with other irregular brightness variations,
are vaguely present in the historical light curve.

Spectra from 2010 March 14-16, reported by C. Buil\footnote{URL: http://www.astrosurf.com/~buil/v407cyg/obs.htm}, 
showed very broad emission lines for all permitted transitions, especially the H Balmer lines, and 
He I, with an ejection velocity $>$2000 km s$^{-1}$, with superimposed very narrow emission lines.   
The surprise that V407 Cyg was previously considered a classical -- although somewhat unusual -- symbiotic was compounded by 
the high-energy detection of both X-rays (XR) and $\gamma$-rays above 100 MeV (Abdo et al. 2010).  
These were interpreted as the signature of $\pi^o$-production by shock-energized protons in the 
RG wind during the first four days of the outburst, when the velocities exceeded 2000 km s$^{-1}$. 

In this first paper we focus on the shock and its development.  In subsequent papers we will 
present the analysis of the circumstellar material illuminated by the photoionization produced 
during the outburst and a line list constructed using high resolution spectra obtained from the Nordic Optical 
Telescope (NOT) on day 22 of the event, before the peak of the XR emission.


\section{Observations}

Our optical observational data set consists of spectra taken between 2010 Mar 24 and 2010 Jul 16 with the Ond\v{r}ejov Observatory Zeiss 2.0 m telescope coud\'e spectrograph  and with the 2.6 m Nordic Optical Telescope (NOT) fiber-optic echelle spectrograph (FIES, program P40-423).   
The Ond\v{r}ejov spectra taken with the SITe005 800x2000 chip were mainly obtained in the vicinity of H$\alpha$,  supplemented by spectra at both bluer and redder wavelengths.  The dispersion of 0.24\AA\ per pixel provided spectral coverage of approximately 500 \AA, and exposure times ranged from 60 s to 6700 s. The FIES spectra were obtained with a dispersion of 0.023\AA\ px$^{-1}$ in high-resolution mode covering the spectral interval from 3635 to 7364\AA\ and 0.035\AA\ px$^{-1}$ in medium-resolution mode covering the spectral interval from 3680 to 7300\AA.  Exposures ranged from 100 s to 3000 s.  Absolute fluxes for the NOT spectra were obtained using the flux standard star BD +28$^o$4211 observed on Jun 3 and Jun 23 at high resolution. Other spectra were not absolutely calibrated.  All NOT spectra were reduced using IRAF, FIESTools, and our own special purpose routines written in IDL.\footnote{ IRAF is distributed by the National Optical Astronomy Observatories, which are operated by the Association of Universities for Research in Astronomy, Inc., under
cooperative agreement with the US National Science Foundation.}  In several instances, contemporaneous spectra between the two observatories allow us to correlate and cross-calibrate the data. 
The journal of our observations is presented in Tables 1a and 1b, which 
provides the observation start time in the systems Universal Time, Julian Date, JD, and $\Delta$T, the approximate time
since maximum optical light (T0). The T0 parameter has been adopted to be 2010 Mar 10.813 = JD 2455266.313 (see section 3.1) for XR analysis.  
For the journal of observations we adopt the value JD 2455266.30 as a guide to 
following the discussion, based on the three nearly coincident reportings of maximum brightness in CBET \#2199.


\begin{table*}
\caption{Table 1a. Nordic Optical Telescope Observations}
\label{table:1a}      
\centering                 

\begin{tabular}{ lclccrl}
\hline
Spectrum ID& Date & UT & JD-240000 & $\Delta$T (days)$^a$&t$_{\rm exp}$ (sec) & Object$^b$\\
\hline
FItc310098 & 2010-04-01 & 05:28:24.8 & 55287.72806 & 21.4 & 1000 & V407 Cyg HR \\
FItd010076 & 2010-04-02 & 05:38:54.7 & 55288.73536 & 22.4 & 125 & V407 Cyg HR \\
FItd010077 & 2010-04-02 & 05:43:11.5 & 55288.73833 & 22.4 & 400 & V407 Cyg HR \\
FItd230110 & 2010-04-24 & 05:11:12.5 & 55310.71612 & 44.4 & 1000 & V407 Cyg HR \\
FItd270157 & 2010-04-28 & 05:30:37.4 & 55314.72960 & 48.4 & 100 & V407 Cyg HR \\
FItd270158 & 2010-04-28 & 05:33:40.3 & 55314.73172 & 48.4 & 1000 & V407 Cyg HR \\
FIte170040 & 2010-05-18 & 04:52:22.9 & 55334.70304 & 68.4 & 4 & V407 Cyg HR \\
FIte170041 & 2010-05-18 & 04:53:48.6 & 55334.70403 & 68.4 & 100 & V407 Cyg HR \\
FIte170042 & 2010-05-18 & 04:56:54.0 & 55334.70615 & 68.4 & 1000 & V407 Cyg HR \\
FIte260044 & 2010-05-27 & 04:38:28.8 & 55343.69339 & 77.4 & 400 & V407 Cyg HR \\
FIte260045 & 2010-05-27 & 04:46:30.5 & 55343.69896 & 77.4 & 3000 & V407 Cyg HR \\
FItf020080 & 2010-06-03 & 04:03:20.1 & 55350.66898 & 84.4 & 200 & V407 Cyg HR \\
FItf020081 & 2010-06-03 & 04:08:01.8 & 55350.67224 & 84.4 & 2564 & V407 Cyg HR \\
FItf020084 & 2010-06-03 & 04:57:26.2 & 55350.70655 & 84.4 & 500 & BD+28d4211 HR\\
FItf150084 & 2010-06-16 & 04:08:29.0 & 55363.67256 & 97.4 & 200 & V407 Cyg HR \\
FItf150085 & 2010-06-16 & 04:13:11.8 & 55363.67583 & 97.4 & 3000 & V407 Cyg HR \\
FItf220081 & 2010-06-23 & 03:10:23.3 & 55370.63221 & 104.3 & 500 & BD+28d4211 MR\\
FItf220083 & 2010-06-23 & 03:27:37.6 & 55370.64419 & 104.3 & 270 & V407 Cyg MR \\
FItf220084 & 2010-06-23 & 03:33:29.5 & 55370.64826 & 104.3 & 2700 & V407 Cyg MR \\
FItg150076 & 2010-07-16 & 00:00:52.9 & 55393.50061 & 127.2 & 270 & V407 Cyg MR \\
FItg150077 & 2010-07-16 & 00:06:44.7 & 55393.50468 & 127.2 & 2700 & V407 Cyg MR \\
\hline
\end{tabular}
\end{table*}
Note: (a) T0 (JD) = 2455266.30; (b) HR = high resolution, MR = moderate resolution
\begin{table*}
\caption{Table 1b. Ond\v{r}ejov Observations}
\label{table:1b}      
\centering                                      
\begin{tabular}{lclccrl}
  \hline
  Spectrum ID & Date & UT & JD - 2400000. & $\Delta$T (days)$^a$ & t$_{\rm exp}$ (sec) & $\lambda_{\rm min}$ (\AA) \\  
\hline
tc230055 & 2010-03-24 & 01:45:46 & 55279.57345 & 13.3 & 1641 & 6258 \\
tc230056 & 2010-03-24 & 02:13:45 & 55279.59288 & 13.3 & 373 & 6258 \\
 tc230058 & 2010-03-24 & 02:24:21 & 55279.60024 & 13.3 & 447 & 6258 \\
 tc230059 & 2010-03-24 & 02:32:24 & 55279.60583 & 13.3 & 389 & 6258 \\
 tc230060 & 2010-03-24 & 02:39:31 & 55279.61078 & 13.3 & 360 & 6258 \\
 tc230065 & 2010-03-24 & 03:07:52 & 55279.63046 & 13.3 & 328 & 8392 \\
 tc230066 & 2010-03-24 & 03:14:45 & 55279.63524 & 13.3 & 1985 & 8392 \\
 td020076 & 2010-04-03 & 01:59:53 & 55289.58325 & 23.3 & 577 & 6258 \\
 td020077 & 2010-04-03 & 02:10:07 & 55289.59036 & 23.3 & 586 & 6258 \\
 td020078 & 2010-04-03 & 02: 20:32 & 55289.59759 & 23.3 & 599 & 6258 \\
 td030034 & 2010-04-04 & 01:33:58 & 55290.56525 & 24.3 & 541 & 6257 \\
 td030035 & 2010-04-04 & 01:43:38 & 55290.57197 & 24.3 & 586 & 6257 \\
 td030036 & 2010-04-04 & 01:54:04 & 55290.57921 & 24.3 & 626 & 6257 \\
 td050019 & 2010-04-06 & 00:35:32 & 55292.52468 & 26.2 & 936 & 6257 \\
 td050022 & 2010-04-06 & 00:58:26 & 55292.54058 & 26.2 & 3759 & 4753 \\
 td060048 & 2010-04-06 & 23:16:59 & 55293.47013 & 27.2 & 1255 & 6257 \\
 td060055 & 2010-04-06 & 23:48:47 & 55293.49221 & 27.2 & 6718 & 4797 \\
 td060063 & 2010-04-07 & 01:51:17 & 55293.57728 & 27.3 & 1996 & 6400 \\
 td070033 & 2010-04-08 & 02:33:25 & 55294.60654 & 28.3 & 60 & 6257 \\
 td070034 & 2010-04-08 & 02:35:58 & 55294.60831 & 28.3 & 300 & 6257 \\
 td070037 & 2010-04-08 & 02:48:24 & 55294.61694 & 28.3 & 60 & 8392 \\
 td070038 & 2010-04-08 & 02:50:52 & 55294.61866 & 28.3 & 2000 & 8392 \\
 td210056 & 2010-04-22 & 02:37:40 & 55308.60949 & 42.3 & 500 & 8395 \\
 td210057 & 2010-04-22 & 02:47:17 & 55308.61617 & 42.3 & 500 & 6259 \\
 td260020 & 2010-04-26 & 23:12:42 & 55313.46715 & 47.2 & 4135 & 6261 \\
 td260021 & 2010-04-27 & 00:25:59 & 55313.51804 & 47.2 & 1106 & 6261 \\
 td280030 & 2010-04-28 & 23:37:32 & 55315.48440 & 49.2 & 800 & 6260 \\
 td280036 & 2010-04-29 & 02:02:32 & 55315.58509 & 49.3 & 1999 & 8396 \\
 tf040013 & 2010-06-04 & 23:06:20 & 55352.46273 & 86.2 & 3000 & 6258 \\
 tf040020 & 2010-06-05 & 00:15:17 & 55352.51061 & 86.2 & 300 & 8394 \\
 tf040022 & 2010-06-05 & 00:25:57 & 55352.51802 & 86.2 & 2400 & 8394 \\
 tf040030 & 2010-06-05 & 01:19:26 & 55352.55516 & 86.3 & 2000 & 7681 \\
 tf060016 & 2010-06-06 & 20:46:29 & 55354.36561 & 88.1 & 4499 & 6260 \\
 tf070017 & 2010-06-07 & 20:35:33 & 55355.35802 & 89.1 & 3000 & 6740 \\
 tf070021 & 2010-06-07 & 21:32:43 & 55355.39772 & 89.1 & 3600 & 7246 \\
 tf070030 & 2010-06-07 & 23:31:08 & 55355.47995 & 89.2 & 2100 & 6258 \\
 tf140016 & 2010-06-14 & 21:24:24 & 55362.39194 & 96.1 & 3600 & 8393 \\
 tf140020 & 2010-06-14 & 22:32:02 & 55362.43891 & 96.1 & 1800 & 6256 \\
 tg120015 & 2010-07-12 & 21:16:22 & 55390.38637 & 124.1 & 3600 & 6259 \\
  \hline
  \end{tabular}
  
  \end{table*}
Note: (a)  $\Delta$T = T - T0, with T0 (JD) = 2455266.30 

\section{Analysis}

\subsection{{\it Swift} Observations: XRT and UVOT}

{\it Swift} (see Gehrels et al., 2004) started observing  V407 
Cyg three days after the peak of the optical light-curve (2010 Mar 10.813 = JD 2455266.313, hereafter denoted as T0), with both the X-ray Telescope (XRT, Burrows et al., 2005) and the UV/Optical Telescope (UVOT, Roming et al., 2005).  Short observations (1 to 2 ksec) were performed every few days until Apr 2 (T0+23.1 days, JD 2455280, the start of our Ond\v{r}ejov observations), at which point the 
exposure time was increased to $\sim$4 ksec 
every other day (with a $\sim$10 ks observation on 2010 Apr 17; T0+37.6 days). 
From 2010 Apr 28 until May 6 (days 48.6-56.3), 3 ksec were obtained every four 
days. From 2010 May 18 (T0+68.7 days), observations of about 1 ksec were taken weekly until 2010 July 2.7.  Data were processed using version 3.5 of the {\it Swift} 
software (released as part of HEASoft 6.8 on 2009 December 03) and the most 
up-to-date calibration files.

Initially the XRT PC mode count rate remained low, $\sim$0.01-0.02 count s$^{-1}$ 
until T0+15.2 days. The rate started to increase at T0+12.4 days, 
reaching a peak of $\sim$0.26 count s$^{-1}$ on T0+30.5 days. The X-rays then began 
a slow decay until around T0+56 days, at which point the count rate stopped 
decreasing, remaining between 0.08-0.09 count s$^{-1}$. The light curve is shown 
in Fig 1.

The UV source was bright and  most of the data were collected in the uvm2 
filter (central wavelength 2246\AA); observations in u and uvw1 were saturated 
and are not shown. The initial data showed a slow decline (approximately 0.02 
mag day$^{-1}$) that steepened to $\sim$0.09 mag day$^{-1}$ around 40 days after peak; 
after T0+56 days the decay once more slowed to 0.05 mag day$^{-1}$. This reduced 
rate of decline in the UV light curve starts at around the same time as the 
XR flux stops declining, a similar feature appears in the AAVSO $V$ band 
light curve at this time.   As we discuss below, this is the same behavior derived 
from the optical emission lines and, in particular, from the third light continuum 
detected using the photospheric absorption lines from the Mira.
  \begin{figure}
   \centering
   \includegraphics[width=8cm]{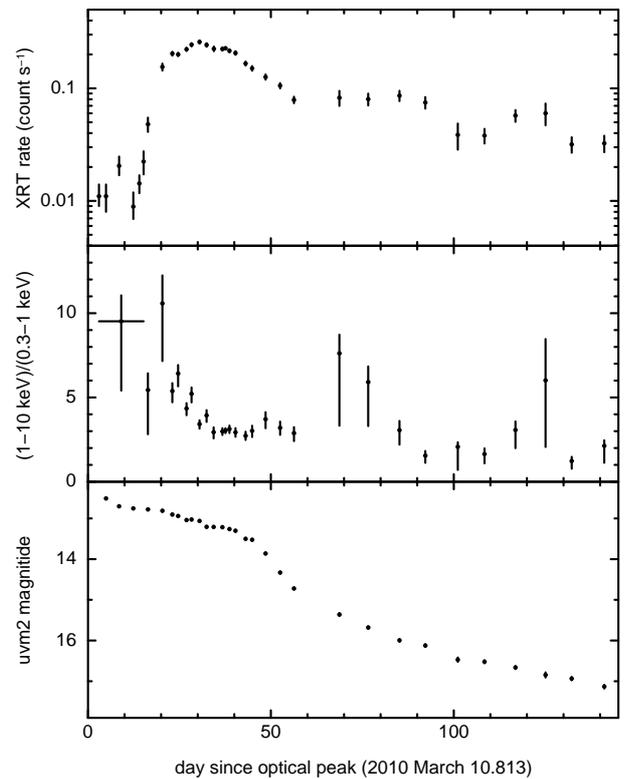}
   \caption{ {\it Swift} XRT 0.3-10 keV count rate, hardness ratio, and UVOT uvm2 
(2246\AA) magnitudes of V407 Cyg as a function of time since the optical peak.   For comparison with the optical spectroscopy, note that T0 =  MJD 55266.}
              \label{outburst-spectra}%
    \end{figure}
    
The XR spectrum remained relatively hard throughout the observations, with 
no strong emergence of super-soft emission. The hardness ratio (defined as 
the count rate ratio [1-10 keV]/[0.3-1 keV]) appears to soften until around 
day 30, after which time it remains approximately constant (see Fig 1).   Spectra were extracted before T0+30 days (as the XR light-curve was brightening), T0=30-60 days (as the X-rays were fading) and after day 60 (when the XR count rate was approximately constant). The earliest two spectra are very flat and cannot be fit with simple models; by $\chi^2$  fitting we ruled out all solar abundance models from single-temperature optically-thin models with a single absorption component to two-temperature models with two absorption components, see Table 2.   We used the Wilms et al. (2000) abundances and absorption model and the mekal plasma emission model. In contrast, non-solar abundance two-temperature models fit these two spectra well. For the late-time (after day 60) spectrum, only a single temperature component was needed; however, a non-solar abundance was still a significant improvement. We were able to achieve a good fit to all three spectra simultaneously, using the same abundances for each, but with other parameters taking different values. While two emission components were required for the spectra before day 60, we could not distinguish between a blackbody and an optically-thin plasma model for the lower temperature component. Both spectral components were absorbed by a single-component cold-absorption model, with the column density decreasing with time. The temperature of the hotter (kT$\approx$2-4 keV) component was also lower in the second spectrum, although the luminosity was unchanged. The (single) temperature of the spectrum extracted after day 60, when the XR count rate remained close to constant, was again lower, while the absorbing column was consistent with the day 30-60 spectrum. The luminosity of the low temperature component is not well constrained due to the combination of its low temperature and high absorbing column. The abundances implied by this fit are very high for N and O, abundances for elements other than those listed were fixed at solar. The simultaneous spectral fit had $\chi^2$/dof = 480/383 (reduced $\chi^2$ = 1.25) for the optically-thin plus blackbody model, and $\chi^2$/dof = 489/385 (reduced $\chi^2$ = 1.27) for the two-temperature optically-thin model. The results of the former fit are shown in Table 3 and Fig 2; all errors and limits are at 90\% confidence. The lower temperature component is responsible for 14\% and 11\% of the counts in the early and middle spectra, respectively. It remains possible that a yet more complex temperature distribution would reduce the implied high N abundance, although we were not able to find a sensible better fit with three temperature components. Consistent with the XRT spectral fits there is no detectable signal in the Swift BAT, with the 15 to 50 keV count rate remaining below 0.003 count cm$^{-2}$ s$^{-1}$.

  For both spectra, the luminosity at 2.7 kpc (Munari et al. 1990) is approximately $1.2\times 10^{34}$ erg s$^{-1}$ for  days 0-30 and 30-60. The day 60+ spectrum has a corresponding luminosity of about half this value,  $6.4\times 10^{33}$ erg s$^{-1}$.  This is the {\it unabsorbed} luminosity, for the {\it hotter} component* only.   The final non-solar abundance enhancements tied between both spectra, with respect to Wilms et al. (2000)  are N  $ >$ 251, 
O   = 35$^{+30}_{-20}$,  Ne  $< $3.1, Na  = 103$^{+103}_{-84}$, and Mg $<$ 2.3.    The upper limit to the Ne and uncertain O abundances do not indicate an ONe white dwarf as the site of the explosion.  They do not, however, exclude it since the ejecta rapidly mix with the environmental material (Walder et al. 2008) the abundances of which are closer to solar than expected from the nucleosynthesis occurring during the nova event.  Thus, the final elemental mixture of the ejecta may be substantially altered during their passage through the Mira wind.

\begin{center}
{\bf Table 2}: $\chi^2$/dof values for solar abundance models for 
the XRT spectra\\
\begin{tabular}{cccc}
\hline
Model         &      &  $\chi^2$/DOF &  \\
 & $<$ day 30   &  day 30-60 & $>$ day 60\\
\hline
mekal*tbabs         &    514/129  &    622/246 & 580/244 \\
mekal*pcfabs         &   263/128    &  615/245 & 580/243 \\
(mekal+mekal)*tbabs   &  210/127  &   476/244 & 430/242 \\
(mekal+mekal)*pcfabs  &  208/126   &   301/243 & 414/241\\
\hline
\end{tabular}
\end{center}

\begin{center}
{\bf Table 3}: Best fit non-solar abundance models for the XRT spectra
\begin{tabular}{ccccc}
\hline
Day &  kT (BB) & kT (mekal) & N$_H$ & \\
after T0 & keV & keV & $10^{22}$ cm$^{-2}$ & \\
\hline
$<$30 & 0.042$\pm$0.003 & 4.10$^{+0.67}_{-0.47}$ & 1.55$^{+0.14}_{-0.15}$ &\\
30-60 & 0.064$^{+0.002}_{-0.004}$ & 1.99$^{+0.13}_{-0.12}$ & 0.62$\pm$0.09 & \\
$>$ 60 & ... & 0.80$^{+0.16}_{-0.12}$ & 0.69$^{+0.11}_{-0.10}$ & \\
\hline\hline
Day & Obs flux & Unabs flux & Obs flux & Unabs\\
after T0 & total & total & hotter & hotter\\
 \hline
$<$30 & 6.17E-12 & 2.44E-8 & 5.81E-12 & 1.09E-11\\
30-60 & 6.10E-12& 6.99e-11& 5.59E-12	& 1.17E-11\\
$>$60 & 1.34E-12 & 6.14E-12	& 1.34E-12 & 6.14E-12\\
\hline
\end{tabular}
\end{center}

%

 \begin{figure}
   \centering
   \includegraphics[width=7cm,angle=-90]{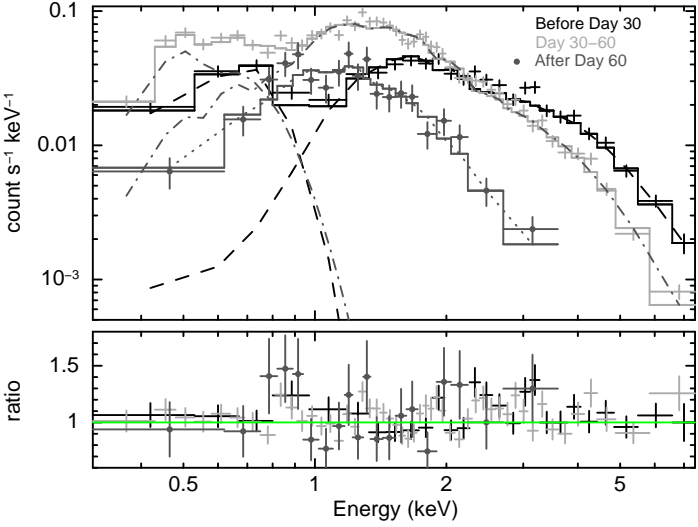}   
   \caption{The joint best fit to the {\it Swift} XRT spectra accumulated before day 30, from day 30 to day 60, and after day 60 from T0.  The model is a strongly absorbed optically thin high temperature component with strongly enhanced nitrogen and oxygen {\it together with a lower temperature blackbody for the first two spectra}.  See text for details. }
              \label{outburst-spectra}%
    \end{figure}

\subsection{Interstellar extinction and spectral type of the Mira}

The analysis of the bolometric properties of the explosion requires a determination of the reddening independent of the XR analysis.  
Using the colors of the Mira variable, and an assigned spectral type of M6 III,  Munari et al. (1990) derived an extinction of $E(B-V)$ = 0.57.  We have instead used optical diffuse interstellar band (DIB) identifications from  Herbig (1975, 1988) and Jenniskens \& Desert (1994).  We used the calibrations for the DIBs given in the latter reference.  Several are either blended with emission lines (e.g. 4428\AA) or too weak to be measured, but most of the important features are accessible.  Our measurements are listed in Table 4, the uncertainty is $\pm$5m\AA.  The resulting reddening is $E(B-V)$=0.45$\pm$0.09 that is consistent with but lower than the value derived from the energy distribution.  We note, however, that the Mira was at minimum light during this outburst.  We have compared the NOT spectra after MJD 55370 with those of giants in the range M6 - M8 taken from the ESO/VLT/UVES {\it Library of High Resolution Spectra of Stars Across the H-R Diagram} archived data set (Bagnulo et al. 2003) and find that the present spectrum  is a far better match to a later spectral type, close to M7 III or M8 III.   In light of the approximate agreement of these different methods, we will adopt $E(B-V)$=0.5 with a range of $\pm 0.05$.    

\begin{center}
{\bf Table 4}: DIBs in the NOT spectrum of V407 Cyg, MJD 55288
\begin{tabular}{ccc}
\hline
Wavelength (\AA) & EW (m\AA) & $E(B-V)$ \\
\hline
5404 & 13  & 0.34 \\
5780 & 234 & 0.49 \\
6162 & 27  & 0.37 \\
6379 & 29  & 0.37\\
6413 & 54  & 0.64 \\
6623 & 109 & 0.47 \\
6843 & 12  & 0.44 \\
\hline
\end{tabular}
\end{center}

Although there are no high resolution ultraviolet (UV) observations of V407 Cyg with which to independently check the interstellar medium along the line of sight, the O star HD 202347 lies at a close Galactic longitude and latitude.  We show in Fig. 3 a comparison of the V407 Cyg Na I D line with interstellar S II 1259\AA\  resonance line in the STIS spectrum HD 202347, a B1 V star in NGC 7039.  This is the closest OB star in Galactic longitude and latitude in the currently available UV archives.\footnote{URL: http://archive.stsci.edu/}  The velocity also agrees with that of the nearby (line of sight)  B8 II star HD 199206, $v_{\rm rad} = -21.0$ km s$^{-1}$ from Wilson (1953) and DDO Radial Velocity Catalog.   This clearly identifies the low velocity component in the V407 Cyg spectrum as interstellar and also shows that the higher velocity components are from the star, especially the wind and circumstellar material.

    \begin{figure}
   \centering
   \includegraphics[width=7cm]{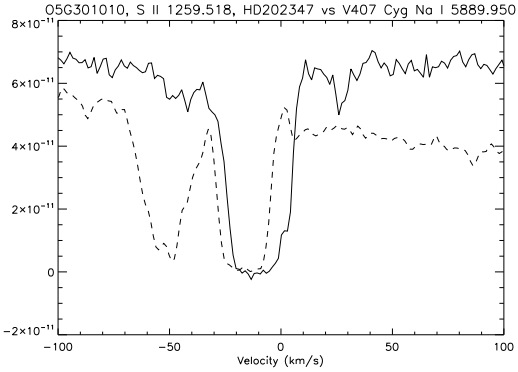}
   \caption{Comparison of the Na I D1 profile of V407 Cyg (dash) with the S II 1257\AA\ resonance line  in HD 202347 (solid, archival {\it HST}/STIS G140H spectrum, program O5G3), along a close line of sight.  The high (negative) radial velocity components are from the Mira wind.  The V407 Cyg spectrum, using the data from the first NOT observation, has been scaled for display; flux units for the $HST$/STIS spectrum are erg s$^{-1}$cm$^{-2}$\AA$^{-1}$}%
    \end{figure}

Several absorption lines, visible in the first NOT spectra, have been used to estimate the continuum contribution of the shock.  These are from low-ionization resonance lines, known to be present in the quiescent spectrum and detected in the spectra of M6-M8 III stars.    We show in Fig. 4 the comparison between the last NOT spectrum and stars from the ESO archive.  The best match is to either M7 III or M8 III.  The reported M6 III spectral classification of Munari et al. (1990) was obtained at a different phase (0.$^p$1, around maximum light) of the Mira while the 2010 outburst occurred at 0.$^p$6 according to the Munari et al. ephemeris.

     \begin{figure}
   \centering
   \includegraphics[width=9.2cm]{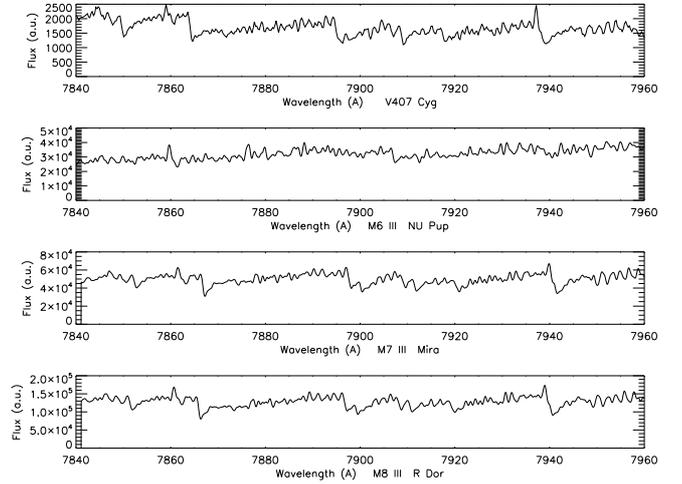}
   \caption{ Comparison of the Ond\v{r}ejov spectrum from MJD 55358 with spectra for M6 III, M7 III, and M8 III stars from the ESO archive.}%
    \end{figure}

\subsection{Low-ionization species}

   \begin{figure*}
   \centering
   \includegraphics[width=12cm,angle=90]{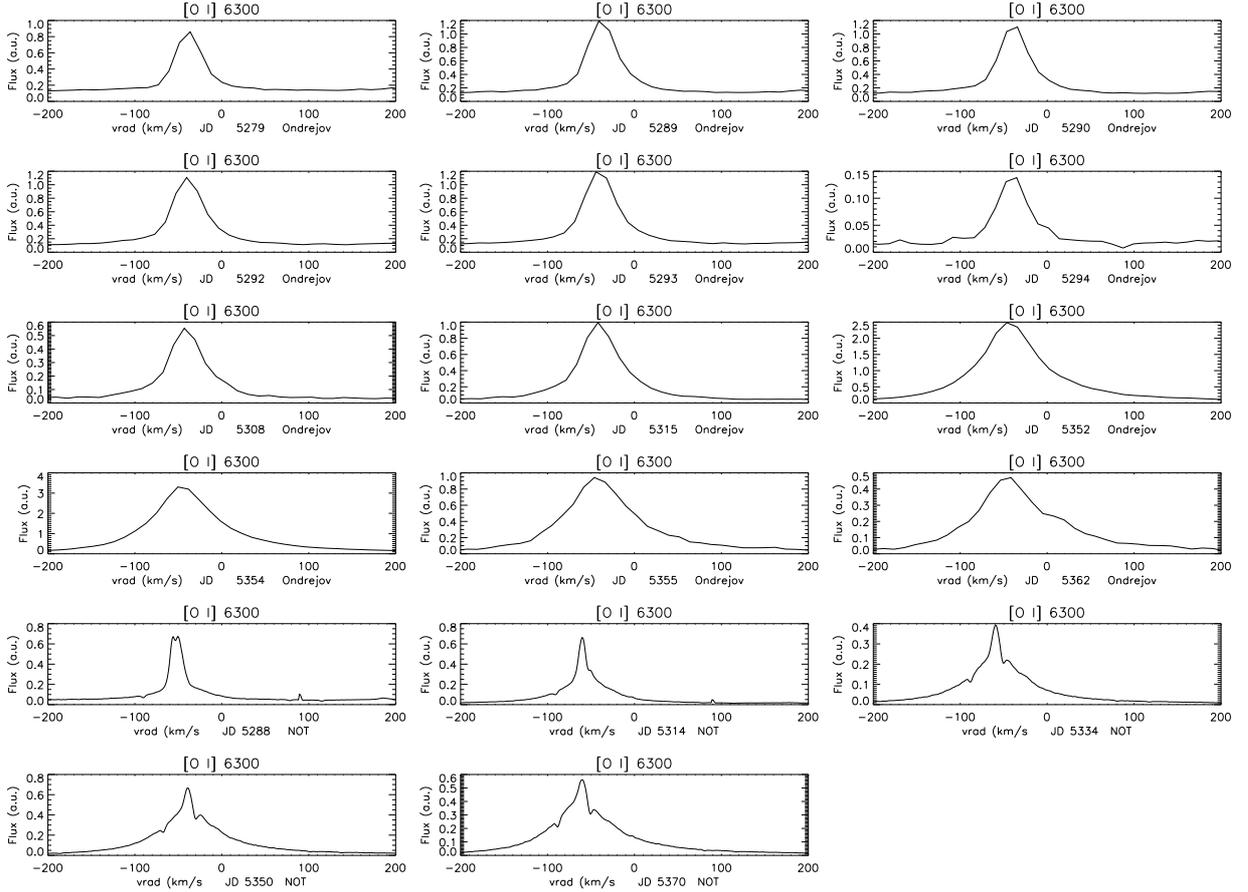}
   \caption{ Variation of the [O I] 6300\AA\ line profile from the Ond\v{r}ejov (11.4 km s$^{-1}$px$^{-1}$) and NOT (1.1 km s$^{-1}$px$^{-1}$) data sets during the observing period.  The date is a JD-2450000; the peak of the XR emission, around day 30, corresponds to JD 2455296. }
              \label{outburst-spectra}%
    \end{figure*}

The O I emission features remained strong throughout the outburst but the line profiles show a complex structure.  In the first NOT observation, the [O I] 6300, 6364\AA\ lines displayed very narrow central emission peaks, FWHM = 12 km s$^{-1}$, with low level, extended wings.  The core emission remained nearly invariant while the broad wings increased throughout the observing interval to a maximum velocity of 200 km s$^{-1}$.   This is shown in Fig. 5  from the NOT and Ond\v{r}ejov sequences.  The FWZI is nearly the same as the broad emission on the Na I D doublet, beginning at around the time of the break in the XR and UV emission seen in Fig. 1.  The O I 8446\AA\ line is fluoresced by Ly$\beta$ through the O I 1027\AA\ resonance line (Bowen pumping), while [O I] 6300\AA\  is produced by recombination and cascade from absorption of the O I 1302\AA\ ground state transition, see Shore \& Wahlgren (2010) for further discussion.   The 8446\AA\ triplet narrowed during the interval from MJD 55279 ($v_{\rm rad,max}$=-900 km s$^{-1}$) to MJD 55315 (-400 km s$^{-1}$) after which it was unobservable in the Ond\v{r}ejov spectra, being overwhelmed by photospheric absorption bands.  We display in Fig. 6 the variations of the equivalent width for [O I] 6300\AA\ and O I 8446\AA.  Note that the peak of the O I 8446\AA\ equivalent width is coincident with the XR peak and the increase of the [O I] 6300 coincides with the break in rate of decline of the UVOT fluxes and the 6700\AA\ continuum.  The maximum velocity observed in the O I 8446\AA\ line is about 500 km s$^{-1}$.  This suggests that the H I Ly$\beta$ pumped transition, O I 1025\AA, was as broad and was at its maximum at the XR peak or that the combined effects of ionization of the environmental Ly$\beta$ and the decrease of the intensity of the O I 1025\AA\  combined to produce the observed variation.  No lines were detected for O II in the NOT spectra.  

The [O III] 4959, 5007\AA\ forbidden lines showed identical profiles and were initially similar to [O III] 4363\AA.  The 4363\AA\ line, however, never developed the double peak structure shown in Fig. 7, displaying only the broad component.  This indicates two separate contributors to the nebular lines, one from material at comparatively large distance that has not been swept up by the shock with a mean density of $10^6$cm$^{-3}$ for T$\approx 10-20$kK, and the other, broader component with a density about a factor of 10 higher in the velocity interval consistent with impact of the ejecta.  
 \begin{figure}
   \centering
   \includegraphics[width=9cm]{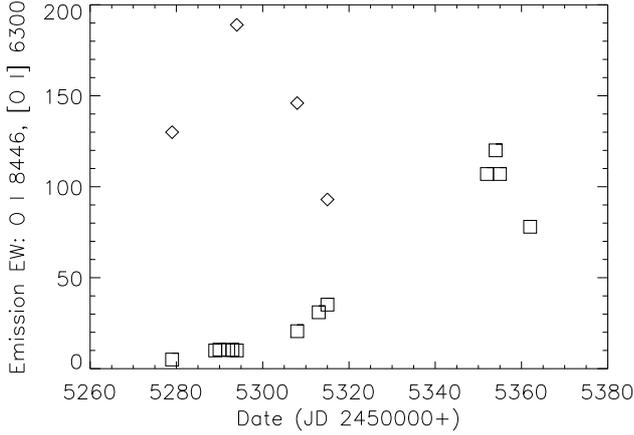}
   \caption{ Variation of the [O I] 6300\AA\ (open squares) and O I 8446\AA\ (open diamonds) equivalent widths during the outburst based on the Ond\v{r}ejov spectra.  }
              \label{outburst-spectra}%
    \end{figure}

    \begin{figure}
   \centering
   \includegraphics[width=9cm]{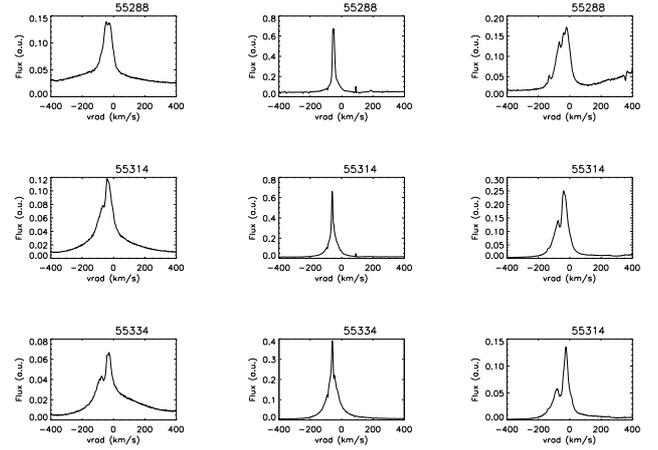}
   \caption{ Comparative development of low ionization species line profiles: [N II] 5754\AA\ (left column), [O I] 6300\AA\ (center), and [O III] 5007\AA\ (right) from NOT spectra on the three indicated dates during the outburst.  }
              \label{outburst-spectra}%
    \end{figure}
    
The [N II] lines, in contrast, show the two narrow components on all three broad profiles (i.e. 5754, 6548, 6583\AA) although the latter two are invisible during the first three NOT observations due to overwhelming blending with H$\alpha$ (see Fig. 7).     They were, however, detectable in the Ond\v{r}ejov spectrum taken on MJD 55308, at the peak of the XR emission.  Their intensities increased relative to H$\alpha$  by about a factor of 5 between MJD 55315 and MJD 55334 and the [N II] (6548+6583)/H$\alpha$ ratio remained nearly constant after MJD 55352.    The equivalent width ratio of [Ni II] 6667\AA\ to He I 6678\AA, which is independent of the extinction and absolute flux calibration, shows a behavior similar to that of the UVOT and 6700\AA\ continuum.   The ratio steadily decreases throughout the outburst.

The [Ar III] 7135.79\AA\ line displayed the same profile as [N II] and [O III]  4959, 5007\AA, with two separate contributing regions.  The photoionized region produced the same comparatively narrow emission peaks and the [O I] 6300\AA\ line has its maximum at the local minimum between the narrow line peaks at -67 and -29 km s$^{-1}$.  The [Ar III] line, however, also developed a broad wing similar to  [Ca V] (see below)   extending from  -200 to +300 km s$^{-1}$ that contributed an approximately equal flux as the narrow lines.   In contrast, [Ar IV] 4711\AA\ showed a  similar profile to that of [Ca V] and He II but  its peak is shifted in all spectra to +110 km s$^{-1}$ (see below, sections 3.4 and 3.6).

The  Ca II IR triplet (8498, 8542, 8662 \AA\ ) lines  all showed broad wings from the earliest Ond\v{r}ejov spectra, extending to 200 km s$^{-1}$ (Fig. 8).  There were several narrower emission peaks in all profiles at the same velocity as the Na I D absorption components.  In contrast, the [Ca II] 7291, 7323\AA\ lines remained extremely narrow, compatible with the [O I] 6300\AA\ core (12 km s$^{-1}$), and never developed the broad wings that are seen on the Ca II 3968\AA\ resonance line or the fluoresced O I lines (Fig. 9).   The Ca II lines showed two very different developments during the outburst.   The 8498, 8542, 8662\AA\ triplet displayed a broad profile in the first observations that systematically narrowed throughout the observing interval, similar to O I 8446\AA.  In contrast, the Ca II 7291, 7323\AA\ doublet remained narrow, $\approx$10 km s$^{-1}$ (FWHM), and centered on the Mira, and never displayed broad wings.
 
    \begin{figure}
   \centering
   \includegraphics[width=7cm]{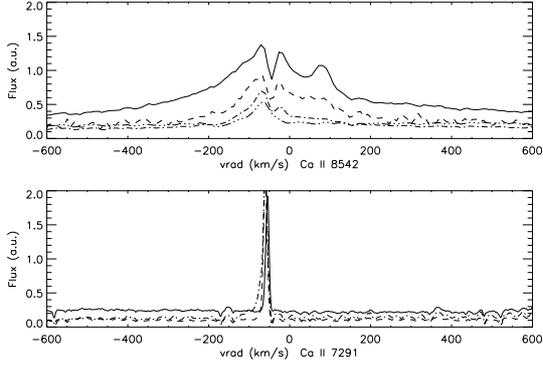}
   \caption{Top: Variation of the main line of the infrared Ca II triplet 8542\AA; solid: MJD 55279, dash: 55294, dot-dash: 55315,  dot-dot-dash: 55362 (Ond\v{r}ejov spectra).  Bottom: Comparison with Ca II 7291; solid: 55289, dash: 55314, dot-dash:55363 (NOT spectra).}
    \end{figure}

 \begin{figure}
   \centering
   \includegraphics[width=7cm]{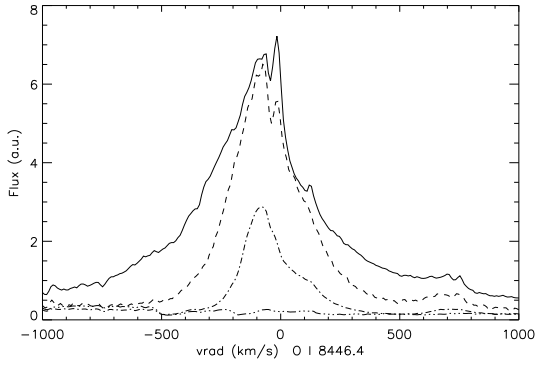}
   \caption{ Variation of  Bowen fluorescence O I 8446 \AA\ line for the same epochs as Fig. 8 (Ond\v{r}ejov spectra). }
    \end{figure}

The Na I D lines displayed a broad component on which were superimposed the narrow absorption features (both ambient and interstellar, see below), extending to 150 km s$^{-1}$.   The broad component increased steadily in strength throughout the observing interval, with the greatest change occurring  after XR maximum.  The centroid of the broad line appears to be shifted by about -16 km s$^{-1}$ relative to the Mira but this may be an artifact of the fitting of a symmetric function (Gaussian) to what may be an intrinsically asymmetric profile of the same form as He II or [Ca V].  In Fig. 10 we show the Gaussian profiles for the two indicated epochs.  The gallery of line profiles in Fig. 24 (see sec. 3.7) shows the changes in the broad component width (along with the narrow circumstellar absorption and emission features).

    \begin{figure}
   \centering
   \includegraphics[width=9cm]{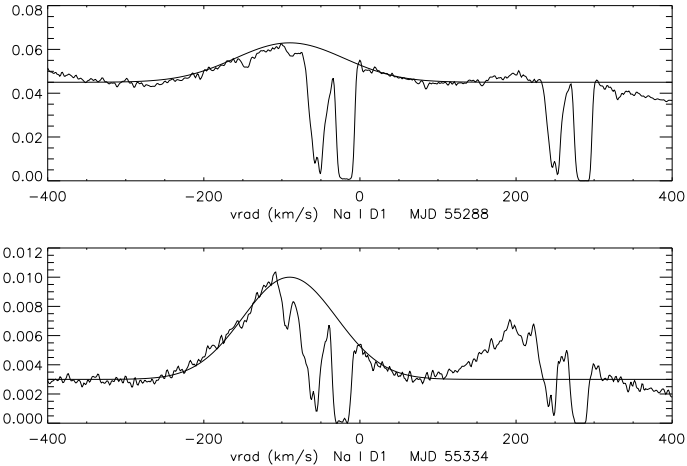}
   \caption{ Gaussian line profiles fit to the Na I D1 profile of the NOT sequence on the dates indicated.  The line FWHM is 110 km s$^{-1}$ for both and the centroid is displaced to  -70 km s$^{-1}$.}
              \label{outburst-spectra}%
    \end{figure}

The Mg I 4571\AA\ profile consisted of two components, like O I and [O I]: a narrow feature coincident with the Mira rest frame and broad wings extending from -200 to +100 km s$^{-1}$ in the first NOT spectra.  The line never displayed a P Cyg component or any of the absorption features observed on the Na I D lines.   It strengthened during the outburst while remaining constant in velocity width.  In Fig. 11 we show the variation of the [O I] 6300\AA\ and Mg I lines corrected for interstellar reddening.  
    
    \begin{figure}
   \centering
   \includegraphics[width=7cm,angle=90]{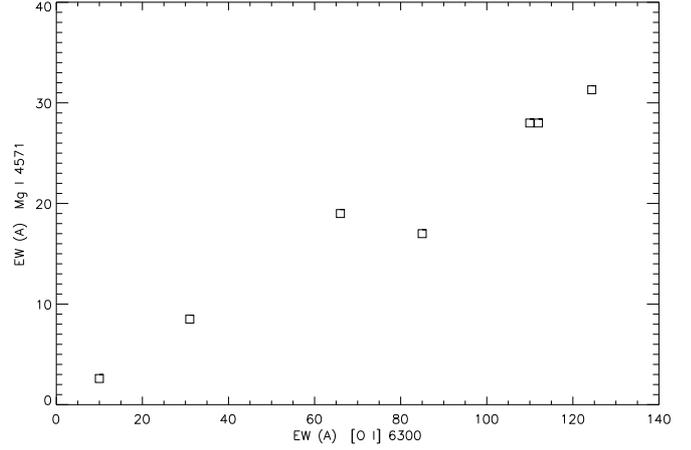}
   \caption{ O I 6300\AA\ equivalent width variations relative to Mg I 4571\AA\ with a reddening correction $E(B-V)$=0.5 applied to the continuum.}
    \end{figure}
The equivalent  variations of the [Ni II] 6667\AA\ line relative to the nearly adjacent He I singlet 6678\AA\ is shown Fig. 12.  This  comparison is unaffected by either reddening uncertainties or continuum variations.   The He I line systematically decreased during the observing interval.  In contrast, the [Ni II] 6667\AA\ line increased in equivalent with showing a change in rate of increase at around MJD 55300 similar to that seen for [O I] 6300\AA\ (e.g. Fig. 6) and [S II] 6730\AA\ (see below, Fig. 26).  
    \begin{figure}
   \centering
   \includegraphics[width=9cm]{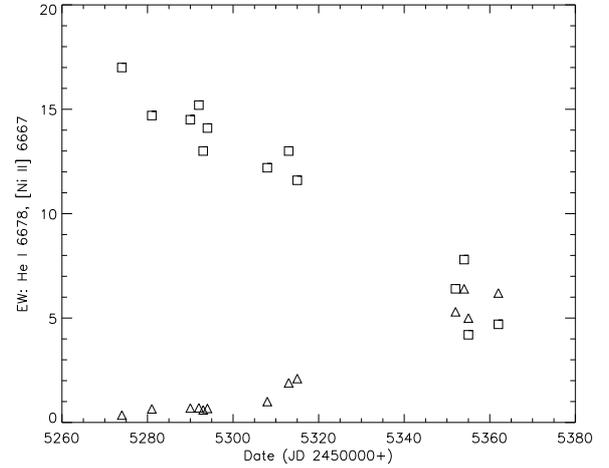}
   \caption{ Equivalent width variation of He I 6678\AA\ singlet (squares) relative to [Ni II] 6667\AA\ (triangles).}%
    \end{figure}

We will present a detailed study of the variations of the Fe-peak neutral and singly-ionized species in the next paper (in preparation).  However, for illustration and comparison with the other neutral and low ionization species, we show in Fig. 13 the P Cyg profiles observed on the narrow components of the Fe II RMT 42 lines observable in the NOT spectra.  The absorption persisted throughout the outburst, even as the wings of the profiles increased in maximum radial velocity to 200 km s$^{-1}$.  The profiles were asymmetric at first, with extended {\it blueward} wings, becoming more symmetric in later spectra.  The broad wings are clearly from the shock while the narrow componwent -- and the P Cyg absorption -- is from the Mira wind and chromosphere.  

    \begin{figure}
   \centering
   \includegraphics[width=9cm]{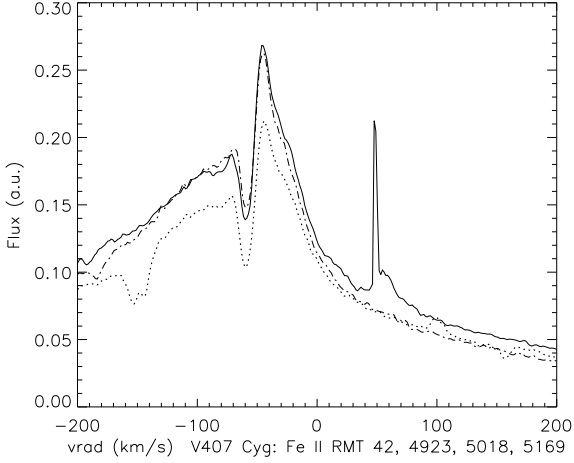}
   \caption{ Low velocity P Cyg absorption on the principal lines of Fe II RMT 42 4923\AA\ (solid), 5018\AA\ (dot-dash), 5169\AA\ (dot) on MJD 55286,  before the peak of the XR emission.  Note also the blueward extension of the line profiles (not shown beyond -200 km s$^{-1}$).}
              \label{outburst-spectra}%
    \end{figure}

The Si II 6347\AA\ line displayed a broad profile throughout the observing interval with wings extending to nearly 400 km s$^{-1}$ and no P Cyg structure at any time.   The companion line at 6371\AA\ was always blended with Fe II 6369.37, 6371.13\AA\ and was unmeasurable.  The profile was very different from the Ca II IR triplet, never showing either the multiple emission features or a strong asymmetry.   The Si II 5041, 5056\AA\  lines   showed broad profiles throughout the observing interval but these are composite, complex blends with [Fe II] 5039, 5043, 5060\AA.  In the first NOT spectrum, however, they are dominated by Si II for which we obtain the  central velocities are both -90$\pm$10 km s$^{-1}$ and FWHM of 300$\pm$18 km s$^{-1}$ from Gaussian fitting.  The lines were nearly of equal strength, and $v_{\rm rad,max}$ was $\pm$400 km s$^{-1}$.  The subsequent line profiles are too complex to permit such a characterization.   Note that the displacement and line width is similar to that of the Na I D broad components.   The line profiles are shown in Figs. 14 and 15.  The comparison to He I 6678 \AA, the only strong singlet observed in the outburst, shows the strong similarity in the profile and evolution.
 \begin{figure}
   \centering
   \includegraphics[width=9cm]{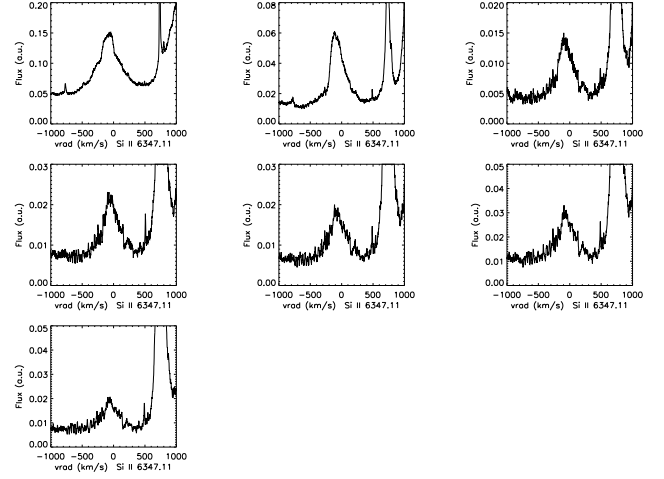}
   \caption{ Variation of  Si II 6347\AA\ line profile in the NOT sequence.  The first spectrum has a contamination from H$\alpha$ that affects adjacent orders from the extreme saturation of the profile.  This changes the core from -150 to +150 km s$^{-1}$ but has no effect on the extended wings.  }
    \end{figure}

 \begin{figure}
   \centering
   \includegraphics[width=7cm,angle=90]{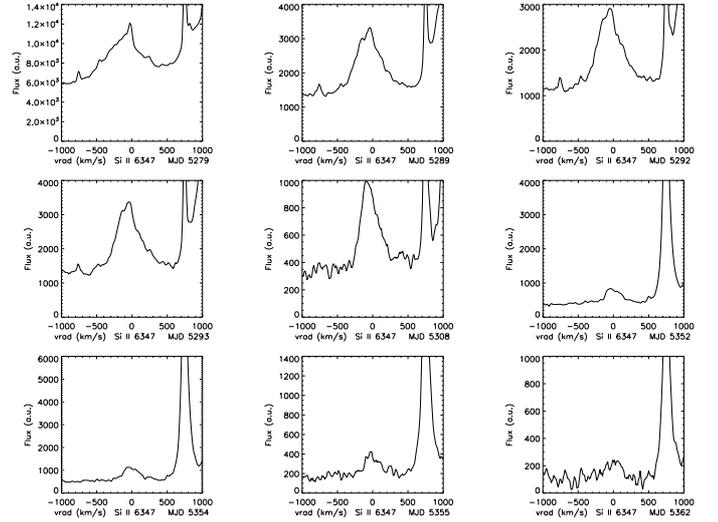}
   \caption{ Variation of  Si II 6347\AA\ in the Ond\v{r}ejov sequence.  Compare this with Fig 17 for He I 6678\AA.  The profiles are unaffected by the contamination present in the NOT echelle spectra at any epoch.}
    \end{figure}

\begin{figure}
   \centering
   \includegraphics[width=7cm]{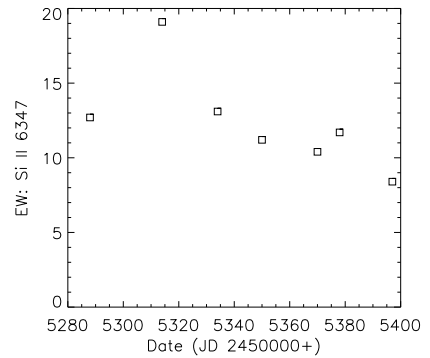}
   \caption{ Variation of  Si II 6347\AA\ equivalent width from the NOT spectra.  The peak in the emission occurs just after XR maximum.}
  
    \end{figure}

The He I line profile variations are shown in Figs. 17 and 18.  Except for 6678\AA, {\it no} singlet line of neutral helium was observed.  The profiles of the 6678\AA\ line, and those of the triplet series, evolved similarly.  The first NOT spectrum displays a broad (nearly $\pm$500 km s$^{-1}$), nearly symmetric profiles for the triplets and 6678\AA, but the subsequent development is different.  The 6678\AA\ profile narrows but remains symmetric while the triplet lines develop multiple narrow emission features and an asymmetric shape extending from -200 to +500 km s$^{-1}$.   The minimum at the peak of the lines corresponds precisely to the maximum of the narrow [O I] 6300\AA\ emission, as found for [N II] and [O III].    

 \begin{figure}
   \centering
   \includegraphics[width=7cm,angle=90]{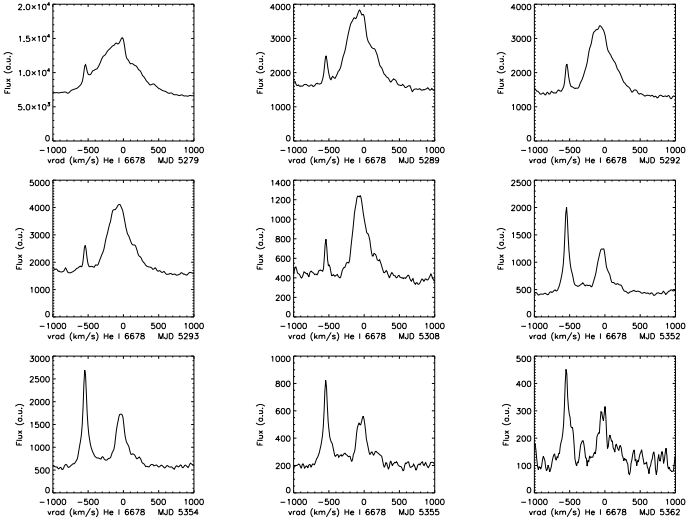}
   \caption{ Variation of  He I 6678\AA\ in the Ond\v{r}ejov sequence for comparison with the Si II 6347\AA\ profile evolution.}
    \end{figure}

The He I line profiles developed very differently than during the 2006 RS Oph outburst.  In the series shown in Fig. 51 of Iijima (2009), in the first week of the RS Oph outburst the profile showed a nearly triangular shape with wings extending to $>$1500 km s$^{-1}$, accompanied by narrower emission peaks at -200 and +100 km s$^{-1}$, that had narrowed to 800 km s$^{-1}$ by day 35.  The subsequent evolution is close to a power law with $v_{\rm rad,max} \approx -800 (t/35)^{-1}$ km s$^{-1}$ and similarly for the red wing.   For V407 Cyg, the He I and He II line profiles varied similarly except that the 5875\AA\ line showed two narrow emission components at the line peak centered on the velocity of the Mira.  The He II profile in RS Oph, from Fig. 50 in Iijima (2009),  displayed multiple components throughout the outburst after day 35 and its broad wings remained symmetric, in contrast to V407 Cyg.     As in V407 Cyg, the singlets were weak or absent in RS Oph 2006.  Only narrow components are reported for any singlet, e.g. 4921\AA.  The He I 6678, 7065\AA\ lines in RS Oph decreased steadily in time while remaining nearly constant relative to H$\beta$, the same was observed for V407 Cyg.

\subsection{H I Balmer line profiles}

The gallery of NOT spectra for the Balmer lines is shown in Fig. 18.  
The maximum velocity reported for the first Balmer line profiles (Mar 14 = MJD 55269, C. Buil, private communication) was $>$2500 km s$^{-1}$ for H$\alpha$ and H$\beta$.   The first Ond\v{r}ejov observation, (Mar 25, MJD 55280) shows, instead, extended emission to 4000 km s$^{-1}$.  This measurement is consistent with the Mar 14 spectrum since the  highest velocity is only seen in very weak wings (the S/N ratio for the first Ond\v{r}ejov H$\alpha$ profile is about 60 pix$^{-1}$ at 4000 km s$^{-1}$ and $>$ 1000 at line peak).    The profiles show a low velocity P Cyg absorption that remains strong throughout the period covered by these observations for the Balmer $\alpha$ through $\gamma$  lines, at -54 km s$^{-1}$.  This is consistent with the rest velocity determined from the Li I 6707\AA\ and Ba II 4554\AA\ resonance lines.  Weaker absorption lines of metallic species seen on the Balmer profiles, especially on H$\beta$,  decreased systematically until MJD 55332, after which they are not detectable.   The H$\delta$ P Cyg trough at low velocity decreased systematically and turned into emission around the date of the UVOT light-curve descent.

 \begin{figure*}
\centering
   \includegraphics[width=12cm,angle=90]{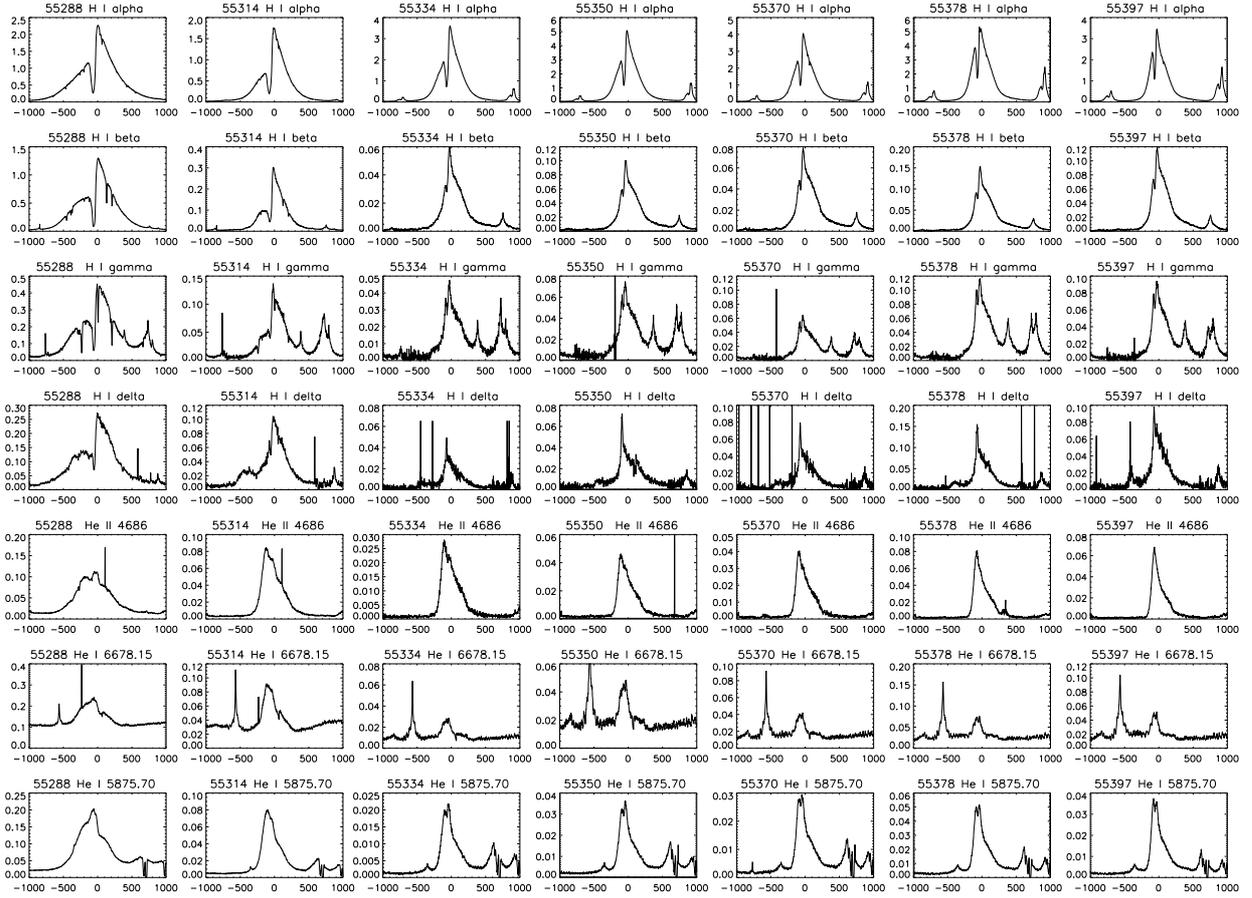}
     \caption{Line profiles for H I Balmer H$\alpha$, H$\beta$, H$\gamma$, and H$\delta$;  He II 4686\AA; He I 6678\AA\ (singlet) and 5875\AA\ (triplet)  for all NOT observations.  The abscissa is uniformly radial velocity in km s$^{-1}$, the individual lines and MJD are indicated in the heading of each panel.  The ordinate is flux in arbitrary units.}
     \label{<Your label>}
\end{figure*}
 
 Figure 19 shows the measured extreme radial velocities for the H$\alpha$, H$\beta$
, and H$\gamma$ profiles.  Blending prevented measurement of the extreme  of the red wing of H$\gamma$ and the low signal-to-noise ratio at H$\delta$ made the measurements unreliable.  The uncertainties are $\pm$20 km s$^{-1}$ based on the continuum fitting.  These should be compared with the He II 4686\AA\ measurements shown below in Fig. 26.

    \begin{figure}
   \centering
   \includegraphics[width=7cm,angle=90]{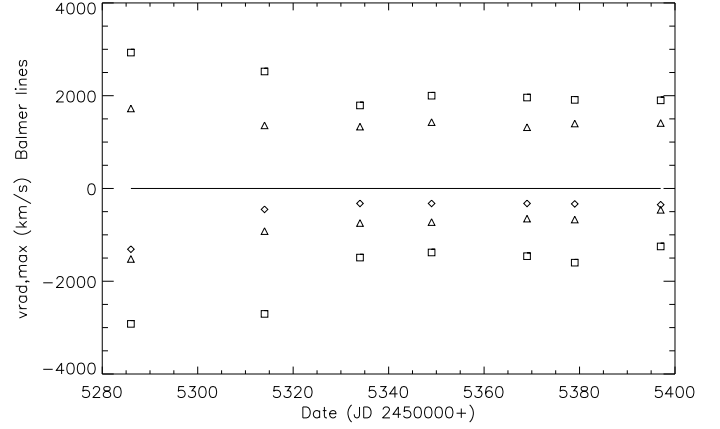}
      \caption{ Maximum radial velocity variation for the Balmer lines from the NOT spectra.  Square: H$\alpha$, triangle: H$\beta$, diamond, H$\gamma$.  }
              \label{outburst-spectra}%
    \end{figure}
 
\subsection{Li I 6707\AA\ and non-photospheric continuum emission}

The Li I 6707\AA\ line was visible from the first spectrum for which we measured an equivalent width, $W_\lambda=42\pm 5$m\AA.  The evolution of the line indicates that it was not formed in the wind but, instead, is the photospheric absorption feature previously reported by Tatarnikova et al. (2003a,b), who reported $W_\lambda = 350$ m\AA.  Its radial velocity, -54 km s$^{-1}$, is consistent with the Mira being at minimum light.  During the first two months, the line steadily increased until, after MJD 55350, it stayed constant.  For the last two NOT spectra (medium resolution; MJD 55370, 55393) we obtained an equivalent width of 350$\pm$10 m\AA.  If we assume the absorption line strength was, in fact, constant throughout the outburst, then the change was due to a ``contamination'' by a masking continuum; thus the additional continuum accounted for $\approx$90\%  of the light at 6700\AA\ during the earliest observation (MJD 55286) and is consistent with the observed decrease of 3 mag in $R$ during the interval.  This is consistent with the simultaneous increase in the atomic and molecular absorption line spectra longward of 6700\AA, with the gradual approach to the pre-outburst spectrum shown in Munari et al. (1990).   The measurements are shown in Fig. 20.   As we will discuss in sec. 4, we interpret this continuum as probably being due to thermal bremsstrahlung emission from the shock.

    \begin{figure}
   \centering
   \includegraphics[width=9cm]{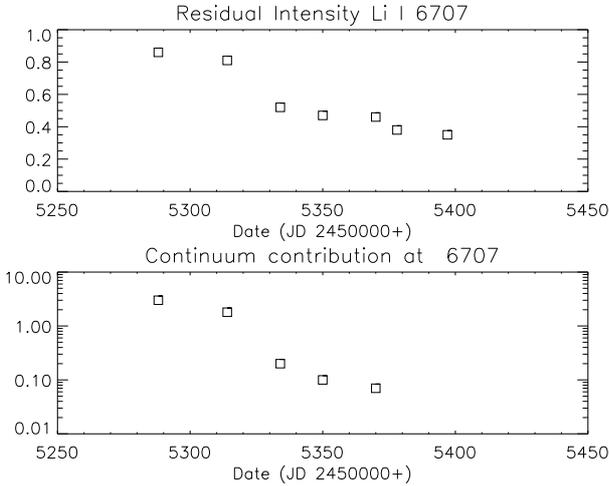}
   \caption{  Li I 6707-derived continuum contribution during the outburst, based on NOT spectra.}
              \label{outburst-spectra}%
    \end{figure}
    
\subsection{High-ionization species}

Shortly after the detection of $\gamma$-ray emission by {\it Fermi}, Munari et al. (2010) reported the presence of high-ionization species in the optical spectrum.  Our NOT sequence, covering the same period at higher resolution, allows us to confirm several of the identifications.  The [Fe X] 6375\AA\ line was present on the first Ond\v{r}ejov spectrum (Mar 25) and continued to increase in strength through May.  After the XR peak, it rapidly declined while maintaining the same line profile, almost identical to He II 4686\AA.  The line, initially nearly symmetric with maximum velocity of $\sim$500 km s$^{-1}$, became progressively more asymmetric in the same manner as the helium lines and virtually disappeared by Jun 2.  The [Ca V] lines, in contrast, remained visible throughout the observing interval and increased in strength through the XR peak, after which they declined less rapidly than [Fe X].  For [Ar X] 5535\AA, the detection is ambiguous because of the presence of a set of strong permitted Fe-peak lines superimposed on a broad but weak emission.  The line is likely present in the first NOT spectrum.

Broad emission was first observed as early as Mar 25.  Munari et al. (2010) report the first detection of [Fe X] on Mar 17, only one week after outburst, with a FWHM of about 600 km s$^{-1}$.  This contrasts with Iijima (2009) who reports,  for the 2006 outburst of RS Oph, that [Fe X] and [Fe XIV] appeared only after day 35.  [Fe X] 6375\AA\ showed a nearly symmetric profile before and at XR maximum, while [Ar IV] 4711\AA\ displayed a He II-like profile with an extended blue wing with a blueward maximum v$_{\rm rad}$=$-$150 km s$^{-1}$ in the stellar reference frame,  [Fe X] 6375\AA\ extended from $-$400 to +600 km s$^{-1}$.  This difference in the profiles persisted, the maximum blueward extension of [Fe X] was always greater than [Ar IV].    The [Ca V]  5309, 6086\AA\ lines were strong and showed a He II-like profile.  These are compared in Figs. 21 and 24.   [Ne IV] 4714, 4720\AA\ and [Fe XIV] 5303\AA\ were absent.  The feature observed near [Ar X] 5533\AA\ is more likely an Fe II permitted line, based on its profile in the first NOT spectrum and no other [Ar X] line was detected.  
    \begin{figure}
   \centering
   \includegraphics[width=8cm]{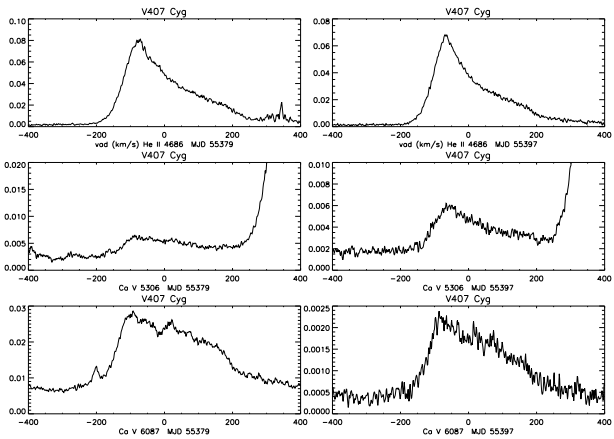}
   \caption{The changes in line profiles for He II 4686 (top), [Ca V] 5308\AA\ (middle), and [Ca V] 6087\AA\ (bottom) on the last two NOT observations.  The similarity of the two sets of profiles demonstrates that the asymmetry is a dynamical effect and not an optical depth effect (see text).}
    \end{figure}
   
Figure 22 shows the [Fe VII] line evolution in the NOT sequence and in  Fig. 23, the last  three epochs are compared to a simple model.  The profile was computed  using a spherical expansion Monte Carlo method described in Shore et al. (1994).  The maximum velocity was fixed to 400 km s$^{-1}$, consistent with the He II line blue wing, and a ballistic velocity law was used.  The only adjustment required is to shift the line centroid to +70 km s$^{-1}$. The best agreement was achieved with a fractional thickness, $\Delta R/R = 0.6\pm0.2$.   The model reproduces the profile best in the two earliest stages, during the initial deceleration of the ejecta, while the last shows an asymmetry on the red wing, likely due to a change in the emission measure of the lower density gas with respect to that of the approaching side of the shock in the denser wind of the Mira.

    \begin{figure}
   \centering
   \includegraphics[width=8cm]{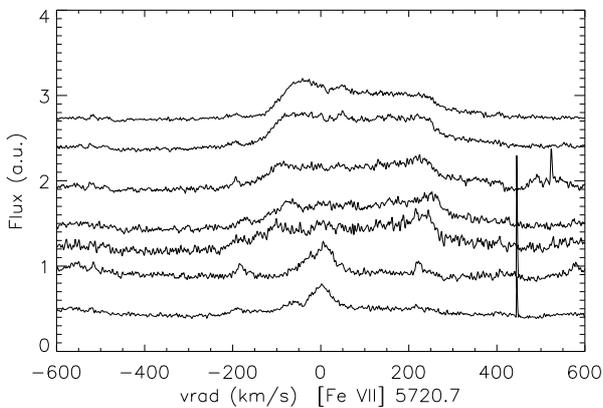}
   \caption{Time development of [Fe VII] 5720\AA\ in the NOT spectra, from bottom to top in the same sequence shown in Fig.  24.}
    \end{figure}

    \begin{figure}
   \centering
   \includegraphics[width=6cm,angle=90]{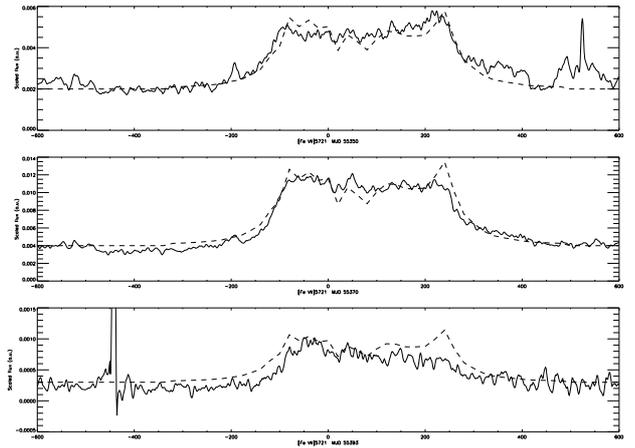}
   \caption{The observed [Fe VII] 5720\AA\ line profiles (from MJD 55363, 55370, 5593) compared with the simulation described in the text.}%
    \end{figure}
    
\subsubsection{O VI Raman line profiles}

The broad emission features 6825\AA\ and 7080\AA\ present in many symbiotics are Raman scattering lines of O VI, formed by the conversion to long wavelength of O VI 1032, 1037\AA\ doublet photons by the nearby Ly$\beta$ 1025\AA\ line (Schmid et al. 1996, Schmid et al. 1999).  
Although the UV flux was several orders of magnitude greater than a normal symbiotic star and the ionization was sufficient to produce the O VI FUV doublet, no O VI Raman emission was detected at either spectral region during the time interval covered by these observations.   This contrasts with the RS Oph 2006 outburst during which the 6825\AA\ line was detected (Brandi et al. 2009, Iijima 2009) in the early stages after outburst (before day 60).  Brandi et al. note that the line was absent in quiescence.  Our non-detection may be due to the orbital phase at which the eruption occurred.   If the WD was at superior conjunction, the part of the Mira wind in which the O VI conversion occurred may have been occulted.   In addition, unlike RS Oph 2006, at the time when recombination yielded sufficient column densities in Ly$\beta$ to  produce the Raman scattering, after Jun 3, the long wavelength region of the optical spectrum was dominated by the photospheric spectrum of the Mira.

\subsection{Emission line flaring: Na I 5889, 5895\AA\ and [S II] 6716, 6730\AA}

The absorption line evolution of the Na I D lines has been discussed recently for the circumstellar environment of several supernovae (Patat et al. 2007, Simon et al. 2009).  We observed a systematic evolution of the absorption of several components during the post-outburst interval.  In the first spectrum (MJD 55287), there is a very weak feature at -95 km s$^{-1}$ with $W_\lambda=5\pm1$m\AA.  Other absorption components were at -63, -59, and -54 km s$^{-1}$.  These latter remained virtually unchanged throughout the entire sequence and we checked the variations relative to the interstellar absorption feature at -23 km s$^{-1}$.  No variations greater than 10\% were measured.  Instead, the -95 km s$^{-1}$ feature systematically grew in equivalent width, reaching a maximum in our last NOT observation (MJD55393) of 120$\pm$2 m\AA.  At the same time, we noted the increase in an {\it emission} line at +6 km s$^{-1}$ that lagged the increase in the absorption by approximately two weeks.   The sequence of line profiles is shown in Fig. 24.  

We propose that this is the {\it rear} side of the same shell that is responsible for the absorption at -95 km s$^{-1}$, and that the delay measures the light-travel time across the structure.  Assuming the measured velocity (in the absence of a dynamical model for the origin of the shell), the delay corresponds to a distance of approximately $4\times 10^{-3}$ pc ($\approx$900 AU) and a tentative time of origin of about one century ago.  This is relatively close to the epoch of the 1936 ``novalike'' event reported by Hoffmeister that has been responsible for the denomination ``nova'' for V407 Cyg in the subsequent literature (a misreading of the catalog entry).  However, it is well known that symbiotic stars undergo outbursts, presumably thermonuclear runaways, during which time the degenerate mass accreter develops a strong wind.  If this happened in the extended outburst in the 1930s, then we may be seeing that ejection through the Na I variations.  The timescale for the increase in the absorption line then suggests an electron density of about 6$\times 10^5$ cm$^{-3}$ at that distance.  Without knowing the thickness of the shell, we can nonetheless estimate the mass-loss rate using the column density, finding for this component $\dot{M} \approx 4\times 10^{-6}$M$_\odot$yr$^{-1}$.  Since the 1936 event lasted nearly a decade, this amounts to a comparatively low mass of about $10^{-4}$M$_\odot$, of which at least some part must be dust.

    \begin{figure}
   \centering
   \includegraphics[width=9cm]{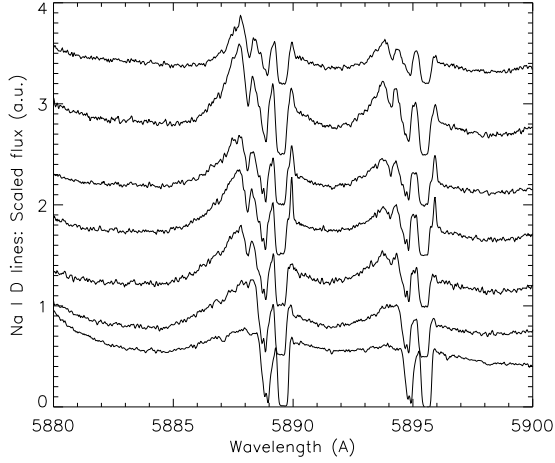}   
   \caption{ Evolution of the Na I D doublet during the outburst (proceeds from bottom to top) from the NOT spectra.  The spectra have been normalized  for display.  Note the appearance of the absorption on both components of the doublet at  5887.9 and 5894.6\AA\ and the narrow emission at 5890.0 and 5896.7\AA.  See text for discussion.}              \label{outburst-spectra}%
    \end{figure}

The [S II] 6716, 6730\AA\  lines displayed a different profile toward the end of the sequence than the first, narrow profiles with a centroid velocity of -60 km s$^{-1}$, hence very low velocity with respect to the star, spanning -150 to +50 km s$^{-1}$.  They remained centered near the stellar velocity and never displayed the redward extension of the higher ionization lines.  A curious feature of their late time evolution is an apparent flare (see Fig. 26) after MJD 55350 during which time the profile did not change.   Flares of other low ionization species have been noted in Iijima (2009) for the RS Oph 2006 event.  This is more consistent with a change in the ionization state of the medium than an impact of the shock.  In fact, it appears that the [S II] 6716, 6730 doublet was not formed in the shock but rather in the Mira, near the base of the wind.  It indicates, however, the interval over which the other high ionization species are ``contaminated'' by environmental emission.  In particular, the peak of the He II line is partly contributed by the low velocity broad emission and this accounts e.g. for the difference between it and the [Ca V] and [Fe X] profiles.     
     
       \begin{figure}
   \centering
   \includegraphics[width=6cm,angle=90]{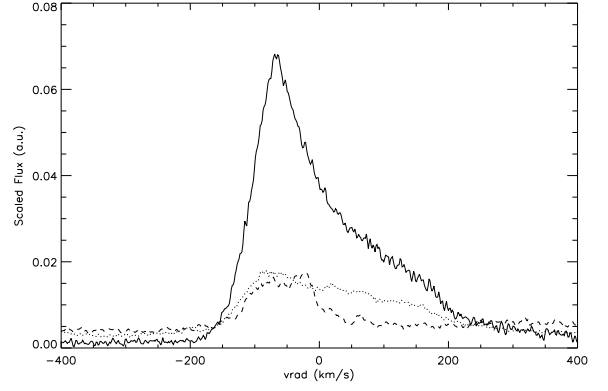}
   \caption{ Comparison between [S II] 6730\AA\ (dash), [Ca V] 6087\AA\ (dot), and He II 4686\AA\ (continuous) line profiles at the last NOT observation, MJD 55397.  Note that the [S II] lacks the redshifted wing present on the higher excitation lines (see text for discussion). }
    \end{figure}  
    
        \begin{figure}
   \centering
   \includegraphics[width=8cm]{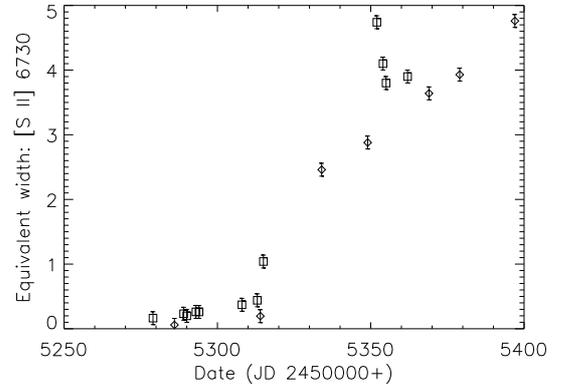}
   \caption{ Variation of the [S II] 6730\AA\ equivalent with showing the shortlived ``flare'' event at about day 84 of the outburst.  There was no profile change during this increase.  Square: Ond\v{r}ejov spectra, diamond: NOT spectra.  Note the rapid increase in line strength approximately coincident with the peak and initial decline of the XR emission (see Fig. 1).}%
    \end{figure}

    \section{Discussion: Toward a model of the outburst and the binary system}

The 1985 and 2006 eruptions of RS Oph are the only symbiotic-like recurrent novae that have been followed throughout the event with modern, multiwavelength methods, and it is that system with which we will compare the 2010 V407 Cyg event.  The optical light curves (see the supplementary material in Abdo et al. 2010) are almost identical although with different timescales, the V407 Cyg event was faster at all wavelengths.    Although the spectroscopic development of the outburst was similar to that of the 1985 and 2006 events in RS Oph (Evans et al. 2008, Iijima 2009) there were some important differences.  The maximum Balmer line velocity reached for V407 Cyg never exceeded 3000 km s$^{-1}$ during the earliest stages of the expansion while those for RS Oph were well in excess of 4000 km s$^{-1}$.  The Balmer line profiles were also systematically different, being more nearly broad Gaussians in V407 Cyg and nearly power laws in RS Oph (Anupama \& Prabhu 1989, Iijima 2009).  While the line structure has been interpreted as individual emission regions, the V407 Cyg variations show that the low velocity P Cyg profile at around -90 km s$^{-1}$ is absorption that changes to emission late in the outburst (particularly evident on the H$\delta$ line).  The persistence of the P Cyg absorption on both the H$\alpha$, H$\beta$, H$\gamma$, and Fe II (permitted) lines at low velocity (around +10 km s$^{-1}$ in the system frame) is similar to that in the RS Oph 1985 and 2006 outbursts.  A difference is the switch to emission at the same wavelength for H$\delta$ at around the peak of the XR emission.  The absorption was at the same velocity as the persistent narrow emission on the O I lines, particularly the forbidden transitions.  

The extra continuum contribution derived using the Li I  profile shows the same variation as the UVOT light curve.  Scaling to the last date, Day 131 of the outburst, on which the contribution is about 10\% of the photospheric continuum, and using the change in the  Ba II 4554\AA\  line give a continuum consistent with bremsstrahlung assuming $E(B-V)$=0.5.   On MJD 55363 (day 104), for which we have a standard star calibration,  the photospheric flux at 6705\AA, near Li I, was $1.6\times 10^{-14}$erg s$^{-1}$cm$^{-2}$\AA$^{-1}$.  At maximum, in the first NOT spectrum, the extra continuum contributed a factor of $\approx$10 times higher (Fig. 20, bottom) at a time when the shock temperature from the XR {\bf mekal} fit was $\approx 2\times 10^7$K, implying a velocity for the shock of $\approx$800 km s$^{-1}$ for the interval from 30 to 60 days after T0.  This is close to the measured maximum velocity on the Balmer and He II 4686\AA\ lines at the epoch of the first two NOT observations.    The shock could then have powered the high ionization emission with the optical continuum being contributed by cooler gas behind the expanding front, consistent with the emission from the lower ionization lines.    This is also sufficient to produce the strong precursor that is required to account for the rapid appearance of the high ionization narrow components.  The decrease in H I column density can be explained by the passage of the shock through the wind and the decrease in neutral absorber due to photoionization (see also Shore et al. 1996).

    \begin{figure}
   \centering
   \includegraphics[width=7cm,angle=90]{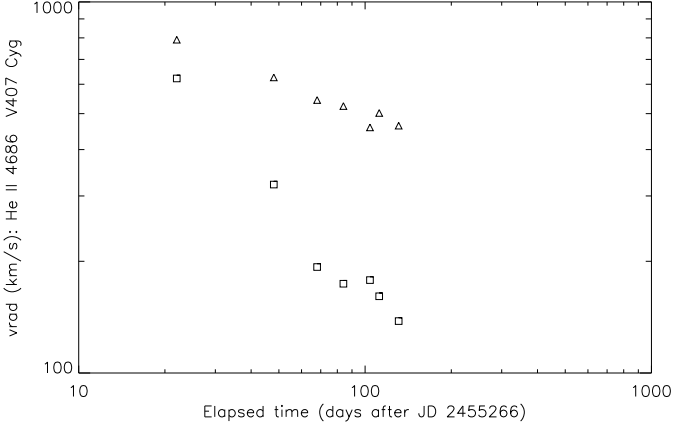}
   \caption{ Maximum positive (triangle) and negative (squares)  radial velocities for He II 4686\AA\ from NOT spectra.  A correction of 50 km s$^{-1}$ has been applied to put the velocities in the rest frame of the red giant  (see text for details).}%
    \end{figure}
    
The changes in the He I, He II, [Ca V], and Balmer line profiles require an aspherical shock, the blueshifted side propagating into the RG wind and the redshifted side displaying breakout of the ejecta.  The terminal velocities can be fit with two power laws, the approaching side with an exponent of -0.84$\pm$0.05 and the receding side with an exponent of -0.31$\pm$.05.  After about day 90, the blue side of the profile changes more slowly and is consistent with the continued expansion of a stalled shock.    Assuming an adiabatic shock in self-similar expansion (Sedov-Taylor) with a power law, and an ambient density varying as $r^{-2}$ with distance $r$ from the Mira (constant mass loss, constant terminal velocity), the blueshifted side that is propagating into the Mira wind follows a velocity scaling as $v\sim t^{-5/7}$ while the redshifted side should vary as $v\sim t^{-1/3}$.  The highest ionization states (i.e. He$^+$, Ca$^{+4}$, Fe$^{+6}$, all of which have ionization potentials $\geq$ 54 eV) display line profiles that are nearly identical among themselves but essentially different from those with lower ionization potentials (i.e. N$^{+}$, O$^{+}$, Ar$^{+2}$, all $\leq$ 54 eV).  In particular, the highest ionization species, Fe$^{+6}$, shows a third form for the line profile and also a systematic redshift.  However, the blueward extension of the lines is asymptotically the same at around -200 km s$^{-1}$ (i.e. $\approx$-150 km s$^{-1}$ relative to the Mira).  There are several possible explanations for the persistent blueward extension of the lines.  One is the wake produced by the convergent shock on the rear side of the RG.  This collimated structure is a generic feature of the explosion and seen in both the Walder et al. (2008) and Orlando et al. (2009) numerical simulations of the RS Oph outburst.  The wake has a far longer lifetime than the shock crossing time and is maintained by the hot post-shocked gas expanding from the cavity generated within the RG wind.  The alternate, more prosaic, explanation is that the approaching side of the shock passed through the denser inner parts of the wind and continued to expand at constant velocity within the increasingly rarefied parts of the RG wind. The Balmer lines were more symmetric and consistent with emission from the entire shock front, while the highest ionization lines show the strongest redward extension and can be explained as the receding portion of the bubble (partly obscured by the Mira) from the low density outer wind.  The lowest ionization species, especially O I and Na I, also displayed asymmetric profiles but with a less pronounced velocity ratio and lower velocities and can be accounted for as the recombintion in the cooling shock.  The virtual absence of H I singlets relative to the triplets, and the broad lines observed on Na I, O I, and Si II in the post-shocked gas may indicate charge exchange effects in a thin shock.  This is supported by the relatively late appearance of the neutral broad lines, at a stage after the asymmetric He II 4686\AA\ line appeared.   Similar effects have been discussed in the context of the Cygnus Loop and other older supernova remnants (e.g. Leutengger et al. 2010).  We will examine this in more detail in the next paper.
    
The similarity of the outburst of V407 Cyg to that of RS Oph is striking, notwithstanding the differences in the properties of the systems.  In this case, modeling performed for RS Oph (e.g. Walder et al. 2008) can also serve for interpreting V407 Cyg.  The explosion should have been non-spherical, although possibly less so than for the much shorter orbital period systems RS Oph and T CrB.  Wind confinement toward the orbital plane may not be so developed in V407 Cyg as in the other two and it is plausible that the ejection was less collimated.  Our simple line profile modeling of [Fe VII] supports this hypothesis.  There is evidently neutral and weakly ionized material that has neither been highly ionized nor swept up in the shock.  However, the disappearance of the narrow absorption lines on H$\beta$ and the changes in the high ionization species' profiles suggests that substantial acceleration has taken place in the ambient matter.  Numerical simulations (Walder et al. 2008) displayed a near doubling of the ejecta mass through mixing with the wind as the shock traverses this ambient medium within the first week.  In the V407 Cyg system this may be longer.  

As to the binary system itself, some quantitative constraints can be placed based on the similarity to RS Oph and the properties of the RG.  The Mira has a long pulsation period and based on the Munari et al. (1990) radius, 400 R$_\odot$, it must not be filling its Roche surface.  If the strong Li I line is evidence of hot bottom burning, as argued by Tatarnikova et al. (2003b) (see also Herwig (2005) for a general review), then its mass must be greater than 4 M$_\odot$.  Assuming that the WD is as massive as in RS Oph, around 1.2 M$_\odot$, then using the scaling from Eggleton (1983) for the ratio of the Roche radius, $R_{RL}$, to the semimajor axis, $a$, gives $R_{\rm RL}/a \approx  0.3$ for a mass ratio of $\ge 3$.  If $R_{RL} > R_{\rm Mira}$, then $a \ge 6$AU, consistent with the {\it Fermi} $\gamma$-ray detection at 3 days after optical outburst (Abdo et al. 2010) for a shock velocity of around 2500 km s$^{-1}$.  The period would therefore be long, as  suggested by Munari et al., (1990), $P_{\rm orb} \ge 32$ yrs.   This is a far longer orbital period than either RS Oph (Brandi et al. 2009) or T CrB, neither of which hosts a pulsating companion to the WD, and supports the conclusion that the degenerates in the symbiotic-like recurrent novae are the most massive among the symbiotic binaries.\footnote{The referee pointed us to a study of RX Pup by Mikolajewska et al. (2002) that finds similarities between that system and both RS Oph and T CrB but with a longer period Mira variable (578 days) and a likely very long orbital period ($>$200 yrs).   The line profile and photometric variations also show symbiotic nova-like outbursts similar to those observed in the 1930's in V407 Cyg.  No explosive ejection has been detected from RX Pup but, in light of the noted similarities, it is worth monitoring.}
    
    \begin{acknowledgements}

PK was supported by ESA PECS grant No 98058. GMW acknowledges support from NASA grant NNG06GJ29G.  APB, JPO \& KLP acknowledge the support of STFC.  We thank  C-C. Cheung,  J. Jos\`e, K. Mukai, U. Munari, Quillo, and C. Rossi for discussions.    We also thank the (anonymous) referee for helpful suggestions.  SNS acknowledges support from the PhD School ``Galileo Galilei'', Univ. of Pisa.  Special thanks J. Mikolajewska for valuable discussions of this and related systems  during her visit to Pisa in May 2010, and the {\it Fermi} LAT group (INFN-Pisa) for collaboration.   Some spectra at Ondrejov were taken by L. Kotkov\'a , P. \v{S}koda, and J. Polster.  
We have made extensive use of the Astrophysics Data System (ADS), SIMBAD (CDS), and the MAST archive (STScI) in the course of this work.
 
\end{acknowledgements}

\end{document}